\documentclass[aps,prd,superscriptaddress,nofootinbib,showpacs]{revtex4}
\usepackage{graphicx}
\usepackage{euscript,amssymb}
\usepackage{amsfonts}
\usepackage{amssymb}
\usepackage[dvips]{color}
\usepackage{color}

\newcommand{\be}{\begin{equation}}
  \newcommand{\ee}{\end{equation}}
\newcommand{\ben}{\begin{eqnarray*}}
  \newcommand{\een}{\end{eqnarray*}}
\newcommand{\bea}{\begin{eqnarray}}
  \newcommand{\eea}{\end{eqnarray}}
\newcommand{\bdm}{\begin{displaymath}}
  \newcommand{\edm}{\end{displaymath}}
\newcommand{\ba}{\begin{align}}
  \newcommand{\ea}{\end{align}}
\newcommand{\lb}{\label}

\newcommand{\arcsinh}{\mathrm{arcsinh}\!}

\renewcommand{\d}{\mathrm{d}\!}
\newcommand{\del}{\partial}

\begin{document}

\title{On the quantum fate of singularities in
        a dark-energy dominated universe}

\author{Mariam Bouhmadi-L\'{o}pez}

\email{mariam.bouhmadi@ist.utl.pt}

\affiliation{Centro Multidisciplinar de Astrof\'{\i}sica - CENTRA,
Departamento de F\'{\i}sica, Instituto Superior T\'ecnico,
Av. Rovisco Pais 1, 1049-001 Lisboa, Portugal}

\author{Claus Kiefer}

\email{kiefer@thp.uni-koeln.de}

\author{Barbara Sandh\"ofer}

\email{bs324@thp.uni-koeln.de}

\affiliation{Institut f\"ur Theoretische Physik, Universit\"{a}t zu
K\"{o}ln, Z\"{u}lpicher Strasse 77, 50937 K\"{o}ln, Germany}

\author{Paulo Vargas Moniz}

\email{pmoniz@ubi.pt}

\affiliation{Departamento de F\'{\i}sica, Universidade da Beira
Interior, Rua Marqu\^{e}s d'Avila e Bolama, 6200 Covilh\~{a},
Portugal}

\affiliation{Centro Multidisciplinar de Astrof\'{\i}sica - CENTRA,
Departamento de F\'{\i}sica, Instituto Superior T\'ecnico,
Av. Rovisco Pais 1, 1049-001 Lisboa, Portugal}

\date{\today}

\begin{abstract}

Classical models for dark energy can exhibit a variety of
singularities, many of which occur for scale factors much bigger
than the Planck length. We address here the issue whether some of
these singularities, the big freeze and the big
\emph{d\'{e}marrage},  can be avoided in quantum cosmology. We use
the framework of quantum geometrodynamics. We restrict our attention
to a class of models whose matter content can be described by a
generalized Chaplygin gas and be represented by a scalar field with
an appropriate potential. Employing the DeWitt criterium that the
wave function be zero at the classical singularity, we show that a
class of solutions to the Wheeler--DeWitt equation fulfilling this
condition can be found. These solutions thus avoid the classical
singularity. We discuss the reasons for the remaining
ambiguity in fixing the solution.

\end{abstract}


\maketitle


\section{Introduction}\label{Intro}

Understanding the observed acceleration of the universe  may affect
 our understanding of  gravity as well as enlarge the framework of
 particle physics, see, for example, \cite{sahnistaro,pad,peebles,FTH}
 for some reviews on the state of art.
 The presence of precision observations is particularly promising in
 this respect \cite{WMAP5}.

Even though a cosmological constant is the simplest way to explain
phenomenologically the late-time acceleration of our Universe, it
is not entirely appealing from a theoretical point of view. The
reason is the mismatch between the required observational value and
the expected one from theoretical grounds \cite{Durrer:2007re}. This
has induced a whole manufacture of theoretical model building,
aiming at an explanation of the recent speed up of the Universe.
Most of these models invoke a dark energy component with negative
pressure \cite{Copeland} or a modification of gravity on large scale
\cite{Durrer:2008in} which, by weakening the gravitational
interaction, achieve the accelerated expansion. More exotic
explanations invoke ideas of a multiverse and may therefore have to
use the anthropic principle or isotropic and inhomogeneous
cosmologies which violate the cosmological principle
\cite{Wiltshire:2007fg,Wiltshire:2007zh,Ellis:2008ar}.

In practical applications and for a homogeneous and isotropic
universe, which is a good approximation for our Universe on large
scales,  whatever the entity responsible for the recent acceleration
may be, it can be described effectively through an equation of state
parameter, $w$. This parameter $w$ is just the ratio between the
pressure and the energy density of the unknown entity (sometimes
designated as ``dark energy'') and may, of course, be time-dependent.
The value of $w$ is observationally very close to $w=-1$, that is,
the equation of state for a cosmological constant. However, if $w$
is larger or
 smaller than $w=-1$, the future of the Universe will be dramatically
 different. Let us mention that indications of a
decaying dark energy have recently been reported \cite{alexei}.

Our main interest here lies not so much in the observational
significance of such models but in their relevance for understanding
the quantization of gravity. The search for a consistent theory of
quantum gravity is among the main open problems in theoretical
physics \cite{OUP}. One aspect is the fate of the singularities
which are prevalent in the classical theory of general relativity.
The hope is that a consistent quantum theory of gravity is free of
such singularities. This aspect can most easily be investigated in
the framework of quantum cosmology: the application of quantum
theory to the Universe as a whole \cite{OUP,CKBS}. Because of our
restriction   to homogeneous models, the corresponding framework of
quantum cosmology does not encounter the mathematical problems of
the full theory. We also emphasize that quantum cosmology is a
non-perturbative framework and thus different from approaches
employing ``quantum corrections'' to the classical
theory\footnote{For reviews on cosmological singularities and
string theory, cf.
\cite{Craps:2007ch,Gasperini:1996fu} 
and references therein.}. Within this context, some of the
dark-energy models are particularly suitable. Furthermore, if one of
the models applies to our real Universe, this would provide us with
important insights into its past and future, which could be a
singularity-free and timeless quantum world.

Indeed, during the last years, it has been shown that some
dark-energy scenarios with $w<-1$, dubbed phantom energy models
\cite{phantom}, can induce a new type of singularity called
big-rip singularity. This is a
singularity where both density and pressure diverge and which is
attained for a universe that expands to infinity in a finite time. The
quantum version of such a situation was extensively
investigated in \cite{DKS} (see also \cite{Elizalde:2004mq,Barboza:2006an}). It was found there that the semiclassical
approximation necessarily breaks down for large-enough scale factors,
and so the singularity theorems no longer apply.

It was soon realized that the big-rip singularity
 is not the unique singularity related to
dark energy and that different types of singularities could show up
\cite{Shtanov:2002ek,Nojiri,sing2}. One of them is the big-brake singularity, which
is obtained for a universe in the future and which is characterized by
an infinite 
deceleration; the universe comes to an abrupt halt. The quantum cosmology
of such a model was discussed at length in \cite{Kamenshchik:2007zj}. 
It was found there that, given reasonable assumptions, the wave
function vanishes in the region of the classical singularity, which we
can safely interpret as singularity avoidance. From theoretical
investigations such classical singularities are well known \cite{Barrowetal},
but in the context here they occur within models possessing
observational relevance.

In this work we are interested in the generalization of these
quantum cosmological results to a broader class of models. We
shall find that the singularity avoidance is there effective, too.
We are mainly concerned with two types of singularities known as the
big-freeze\footnote{It has been recently shown that a big-freeze
singularity might be simply an indication that a brane is about to
change from Lorentzian to Euclidean signature \cite{Mars:2007wy},
even though the brane and the bulk remain fully regular everywhere.
This kind of behavior can happen in some braneworld models where the
bulk is always Lorentzian and does not change its signature. If the
brane and the bulk change simultaneously their signature like in the
models discussed in \cite{Garriga:1999bq,BouhmadiLopez:2002mc},
observers at the brane would not perceive to go into such a
big-freeze singularity.} and big-\emph{d\'{e}marrage} singularity,
respectively \cite{BouhmadiLopez:2006fu}. Let us give a brief
characterization of them:

\begin{itemize}

\item The big-freeze singularity (or type-III singularity in the
  nomenclature of \cite{Nojiri}) takes place at finite scale factor
and finite cosmic time in a (flat)
Friedmann--Lema\^{\i}tre--Robertson--Walker (FLRW) universe. At this
singularity, both the Hubble rate and its cosmic derivative blow up. 
(This is not the case for a big-brake singularity.) Such
a singularity can be induced by a generalized Chaplygin gas (GCG)
\cite{BouhmadiLopez:2006fu}. The GCG is a perfect fluid which
satisfies the following polytropic equation of state \cite{Kamenshchik}:

\begin{equation}\label{eqstate}
p=-\frac{A}{\rho^{\beta}}\ ,
\end{equation}
where $A$ and $\beta$ are constants.  This equation of state was introduced in
cosmology with the intention to unify the dark sectors of the
Universe, that is, dark matter and dark energy \cite{Kamenshchik}. How
this is achieved can be seen from the
conservation of the energy--momentum
tensor of such a fluid in a homogeneous and isotropic universe. It
implies

\begin{equation}\label{dos}
\rho=\left(A+\frac{B}{a^{3(1+\beta)}}\right)^{\frac{1}{1+\beta}},
\end{equation}
where $B$ is a constant. Therefore, if $A, B$, and $1+\beta$ are positive, the
energy density $\rho$ interpolates between dust energy density for
small scale factor and
a constant energy density for large scale factor.
However, it is possible that the GCG can exclusively correspond to
dark energy in a FLRW universe.

The behavior of a GCG can be quite different if the parameters $A,
B$ and $1+\beta$ are not all positive. In particular, it can induce
different sorts of singularities
\cite{BouhmadiLopez:2007qb}. Moreover, it was shown  that a
phantom GCG, that is, a fluid which satisfies the polytropic equation of
state (\ref{eqstate}) for $\rho>0$ and $p+\rho<0$, can induce a
future big-freeze singularity \cite{BouhmadiLopez:2006fu}. It was
also realized that even a GCG fulfilling  the null, strong, and weak
energy conditions can lead to a big-freeze singularity
\cite{BouhmadiLopez:2007qb} in the past.

\item In Ref.~\cite{BouhmadiLopez:2007qb}, it was also shown that a
sudden singularity (or type-II singularity in the nomenclature
of \cite{Nojiri}) can be induced by a
GCG. A sudden singularity is characterized by the fact that the Hubble
rate is finite,  while its
cosmic derivative blows up at finite scale factor.  If the GCG fulfils
the null, strong, and
weak energy conditions, then the sudden singularity corresponds to a
big-brake singularity \cite{Kamenshchik:2007zj} which takes place in
the future when the universal expansion is stopped by an
infinite deceleration. On the other hand, if the GCG
corresponds to a phantom fluid, the sudden singularity takes place
in the past. In analogy with the terminology of the big-brake
singularity we call this kind of sudden singularity a
 big-\textit{d\'emarrage} singularity because the universe starts its
expansion with an infinite acceleration. Since the big-brake singularity
has been studied in \cite{Kamenshchik:2007zj}, we shall
restrict our attention here to the phantom model exhibiting a
big-\textit{d\'emarrage} singularity.
\end{itemize}

An interesting feature of these singularities is the fact that they
can occur in a macroscopic universe, that is, for large values of
the scale factor. A pertinent question is then whether  quantum
gravitational effects can resolve these singularities. Would that
mean that there could be quantum effects in the macroscopic
universe? \emph{How}  could we expect these singularities to be
resolved through quantum gravity? This is the major motivation
standing behind the quantum analysis of the big-freeze and big-{\it
d\'emarrage} singularities. We shall carry out our quantization in
the geometrodynamical framework, using the three-metric and its
conjugate momentum as fundamental variables. The governing equation
in this framework is the Wheeler--DeWitt equation. One can invoke
various reasons why this framework is appropriate for studying the
question at hand\footnote{For a discussion on singularity avoidance
within loop quantum cosmology, cf. \cite{Bojowald:2007ky}; for an
earlier discussion where geometrodynamical elements were present,
cf.
\cite{Berger:1982rd,Berger:1993ff,Berger:1998us,Berger:1999tp}.},
\cite{honnef08}. Perhaps the most compelling one is the fact that
the Wheeler--DeWitt equation is the wave equation which
straightforwardly leads to the Einstein equations in the
semiclassical limit.

Our paper is organized as follows. In Sec.~\ref{2} we
review the classical cosmology with a
generalized Chaplygin gas leading to a past/future big-freeze
singularity or a past sudden singularity.  To be able to study the
quantum behavior, the GCG has to be mimicked by a fundamental
field because also the matter part should have its own degrees of
freedom.
 The representation of the GCG in terms of minimally coupled
scalar fields will be given in Sec.~\ref{3}.
The remainder of this article is concerned with the quantum
properties of these scenarios. In Sec.
\ref{QC}  we obtain solutions for the quantum cosmological model
with phantom and non-phantom GCG evolving to and from a big-freeze
singularity, respectively. We investigate whether this
singularity can be quantum gravitationally avoided. 
The big-\emph{d\'{e}marrage} singularity is also addressed
at the quantum cosmological level. The (general) results are then
discussed for different types of boundary conditions in Sec.
\ref{6}. We present explicit results for the quantum states that
avoid the singularities and discuss further consequences.
Sec. \ref{end} gives our conclusions and outlook. In the appendix, we
present a justification of the gravitational wave function
approximation used in the paper.


\section{The classical big-freeze and big  d\'emarrage singularities
  induced by a generalized
Chaplygin gas}\label{2}
\subsection{The big-freeze singularity without phantom matter}
\label{bfnonphantom}

\begin{figure}[h]
\begin{center}
\includegraphics[width=7cm]{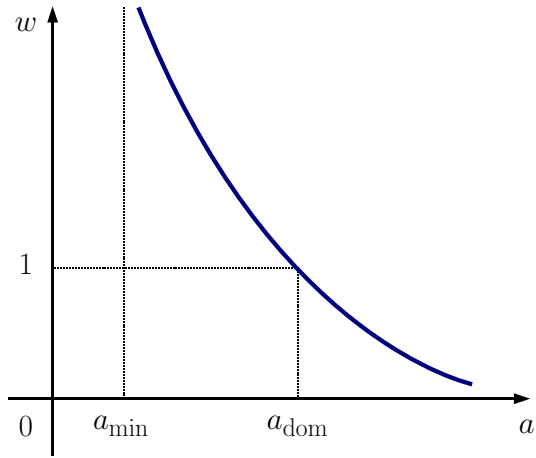}
\end{center}
\caption{Plot of the ratio of pressure and energy density, $w$,
for the generalized Chaplygin gas analyzed in this section as a
function of the scale factor \cite{BouhmadiLopez:2007qb}. The dominant
energy condition is not
fulfilled for a scale factor smaller than $a_{\rm{dom}}$.}
\label{adomfig}
\end{figure}

We review here a particular case of the plain generalized Chaplygin
gas \cite{BouhmadiLopez:2007qb}, that is, a fluid which satisfies
the polytropic equation of state (\ref{eqstate}) and fulfils the
null, strong, and weak energy conditions. Such an equation of state
may induce a big-freeze singularity in the past. This is the
case\footnote{The conclusion reached in this section also holds for
$A>0$, $B<0$, $1+\beta<0$ and $1+\beta=1/(2n)$, with $n$ some
negative integer number. For simplicity, we  shall disregard this
case.} when $A<0$, $B>0$ and $1+\beta<0$. Then the energy density
can be written as
\be
\rho=|A|^{\frac{1}{1+\beta}}\left[-1+\left(\frac{a_{\rm{min}}}{a}\right)^{3(1+\beta)}\right]^{\frac{1}{1+\beta}} 
\ ,
\ee
where
\begin{equation}
  a_{\rm{min}}=\left|\frac{B}{A}\right|^{\frac{1}{3(1+\beta)}}
  \label{amin}
\end{equation}
denotes the minimal scale factor, which is the value where the
singularity occurs. We consider a spatially flat homogeneous and isotropic universe filled
with this sort of fluid. Then, at the minimum scale factor
$a_{\rm{min}}$, the energy density blows up and so does the Hubble
rate. Similarly the pressure, which reads 
\begin{equation}
  p=|A|^{\frac{1}{1+\beta}}\left[-1+\left(\frac{a_{\rm{min}}}{a}\right)^{3(1+\beta)}\right]^{-\frac{\beta}{1+\beta}},
\end{equation}
diverges when the scale factor approaches its minimum value
$a_{\rm{min}}$. Notice that the pressure is positive and therefore a
FLRW universe filled with this fluid would never accelerate.
This particular choice of parameters can thus not describe the
acceleration of the current Universe.
In fact, the deceleration parameter
\begin{equation}
  q=\frac12 (1+3w)
\end{equation}
is always positive.
Figure~\ref{adomfig} displays $w=p/\rho$.
 On the other hand, the Raychaudhuri equation implies that at
$a_{\rm{min}}$ the cosmic-time derivative of the Hubble rate also
diverges.
In order to show that the event that takes place at $a_{\rm{min}}$
corresponds to a past big-freeze singularity, it remains to be proven
that from a given finite scale factor the cosmic time elapsed since
the singularity took place is finite. This can be done by integrating
the Friedmann equation. The cosmic time in terms of the scale factor
reads \cite{BouhmadiLopez:2007qb}
\begin{equation}
  t-t_{\rm{min}}=-\frac{2}{\kappa\sqrt{3}}\frac{|A|^{-\frac{1}{2(1+\beta)}}}{1+2\beta}\left[\left(\frac{a_{\rm{min}}}{a}\right)^{3(1+\beta)}-1\right]^{\frac{1+2\beta}{2(1+\beta)}}\left(\frac{a}{a_{\rm{min}}}\right)^{3(1+\beta)}
  F\left(1,1;\frac{3+4\beta}{2(1+\beta)};1-\left(\frac{a}{a_{\rm{min}}}\right)^{3(1+\beta)}\right), 
\end{equation}
where $\kappa^2=8\pi G$, and
${\rm{F}}(b,c;d;e)$ is a hypergeometric function, see, for example,
\cite{Gradshteyn}. In
the previous expression, $t_{\rm{min}}$ corresponds to the cosmic time
when a universe filled by this sort of generalized Chaplygin gas would
emerge from a past big-freeze singularity at $a=a_{\rm min}$.
 We  thus find here a
singularity at finite scale factor, $a_{\rm{min}}$, and finite cosmic
time, $t_{\rm{min}}$, where both, energy density and pressure, blow
up, as the Hubble parameter and its cosmic derivative do as well. It
can easily be checked that the cosmic time elapsed since the universe
emerged from the past big-freeze singularity until it has a given size
(at a given $t$) is finite\footnote{A hypergeometric series
$\textrm{F}(b,c;d;e)$, also called a hypergeometric function,
converges at any value $e$ such that $|e|\leq 1$ whenever
$b+c-d<0$. However, if $0 \leq b+c-d < 1$ the series does not
converge at $e=1$. In addition, if $1 \leq b+c-d$, the
hypergeometric function blows up at $|e|=1$
\cite{Gradshteyn}.\label{series}}, that is, $ t-t_{\rm{min}}$ is bounded
for any finite scale factor.
Before concluding this section, we notice that although the
generalized Chaplygin gas analyzed here fulfils the null, strong and
weak energy conditions, it violates the dominant energy condition for
scale factors smaller than $a_{\rm{dom}}$, where
\begin{equation}
  a_{\rm{dom}}=2^{-\frac{1}{3(1+\beta)}}a_{\rm{min}}.
\end{equation}
At $a_{\rm{dom}}$ the pressure equals the energy density and for
smaller scale factors $\rho<p$. This situation is schematically shown
in Figure~\ref{adomfig}.

\subsection{The big-freeze singularity with a phantom GCG}
\label{bfphantom}

The big-freeze singularity discussed above may also take place in the
future at a scale factor $a=a_{\rm max}$.
This is possible
if the GCG exhibits a phantom behavior, that is, if it
satisfies $p+\rho<0$ \cite{bl-jm}.
Eventually, the pressure will be  so negative
that the universe does not only accelerate but is even
super-accelerating, that is, the Hubble rate grows as the universe
expands. This is certainly a model capable of describing dark energy.
When the scale factor approaches $a_{\rm{max}}$, the energy
density as well as the pressure diverge in  a finite future cosmic
time \cite{BouhmadiLopez:2007qb}. This phantom GCG  induces a
big-freeze singularity, although in this case the event happens
in the future instead of, as above, in the past
\cite{BouhmadiLopez:2006fu,BouhmadiLopez:2007qb}. Such a future
big-freeze occurs for $A>0$, $B<0$ and $1+\beta<0$. Being more
precise, the energy density in this case reads

\begin{equation}
  \rho=A^{\frac{1}{1+\beta}}\left[1-\left(\frac{a_{\rm{max}}}{a}\right)^{3(1+\beta)}\right]^{\frac{1}{1+\beta}},
  \label{rhobfp}
\end{equation}
while the pressure is
\begin{equation}
  p=-A^{\frac{1}{1+\beta}}\left[1-\left(\frac{a_{\rm{max}}}{a}\right)^{3(1+\beta)}\right]^{-\frac{\beta}{1+\beta}},
  \label{pbfp}
\end{equation}
where
\begin{equation}
  a_{\rm{max}}=\left|\frac{B}{A}\right|^{\frac{1}{3(1+\beta)}}
\end{equation}
is the maximal value of the scale factor.

\subsection{The big-\textit{d\'emarrage} singularity with a
  phantom GCG}\label{bigdemarrage}

The GCG is also known to induce future or past sudden singularities
\cite{Barrow:2004xh,Kamenshchik:2007zj,BouhmadiLopez:2007qb}. The
following cases can be distinguished:
\begin{enumerate}
\item If the GCG fulfils the null, strong, and weak energy conditions
  with $A<0$, $B>0$ and $\beta>0$, a universe filled with this sort of
  gas would face a future sudden singularity, which is just the big-brake
  singularity already mentioned above.
  For $\beta =1$ this fluid was named an
  anti-Chaplygin gas in \cite{Kamenshchik:2007zj} because the
  case with $\beta =1$ and $A>0$ is usually called Chaplygin case.
  Since this model was discussed at length in
  \cite{Kamenshchik:2007zj}, we shall disregard it here.

\item If the GCG is a phantom fluid with $A>0$, $B<0$ and $\beta>0$, a
  universe filled with this fluid would face a past sudden
  singularity, that is, the classical evolution starts at a finite scale
  factor and energy density while the pressure of the fluid diverges.
 We name this past sudden singularity a big-\textit{d\'emarrage}
 singularity as the universe would start its
  classical evolution with an extremely large acceleration due to the
  very negative pressure of the fluid. This phantom GCG corresponds to
  an example where a future big-rip singularity can be avoided even
  for phantom matter \cite{bl-jm} as in this case the universe would
  asymptotically approach de~Sitter space in the future. Here we are rather
  interested in the early behavior of a universe close to the
  big-\textit{d\'emarrage} singularity.
\end{enumerate}

Let us add a few more comments on  the second case stated above. The
energy density and pressure of the phantom GCG can be expressed as
in (\ref{rhobfp}) and (\ref{pbfp}) after substituting
$a_{\rm{max}}$ by $a_{\rm{min}}$ where the initial scale factor
$a_{\rm{min}}$ can be expressed as in (\ref{amin}). The phantom
nature of this fluid induces a super-acceleration of the universe
which in particular implies that a universe filled with this fluid
would be accelerating.


\section{The classical big-freeze and big-\textit{d\'emarrage}
  singularities induced by  scalar fields}\label{3}

A perfect fluid is an effective description of matter. As such it is
valid only on certain (large) scales.
At the quantum
level, we may need a more fundamental description of matter. We
herein choose the simplest possible one: a minimally coupled scalar
field. In the following subsections, we formulate the standard and
phantom GCG in terms of standard and phantom scalar fields,
respectively. This step is an important preparation for the quantum
part because the wave function is defined on configurations space,
that is, it will depend on the scale factor and the scalar field.

\subsection{The big-freeze singularity driven by a standard canonical
scalar field}
\label{bfnonphantomscalar}

\begin{figure}[t]
\begin{center}
\includegraphics[width=8cm]{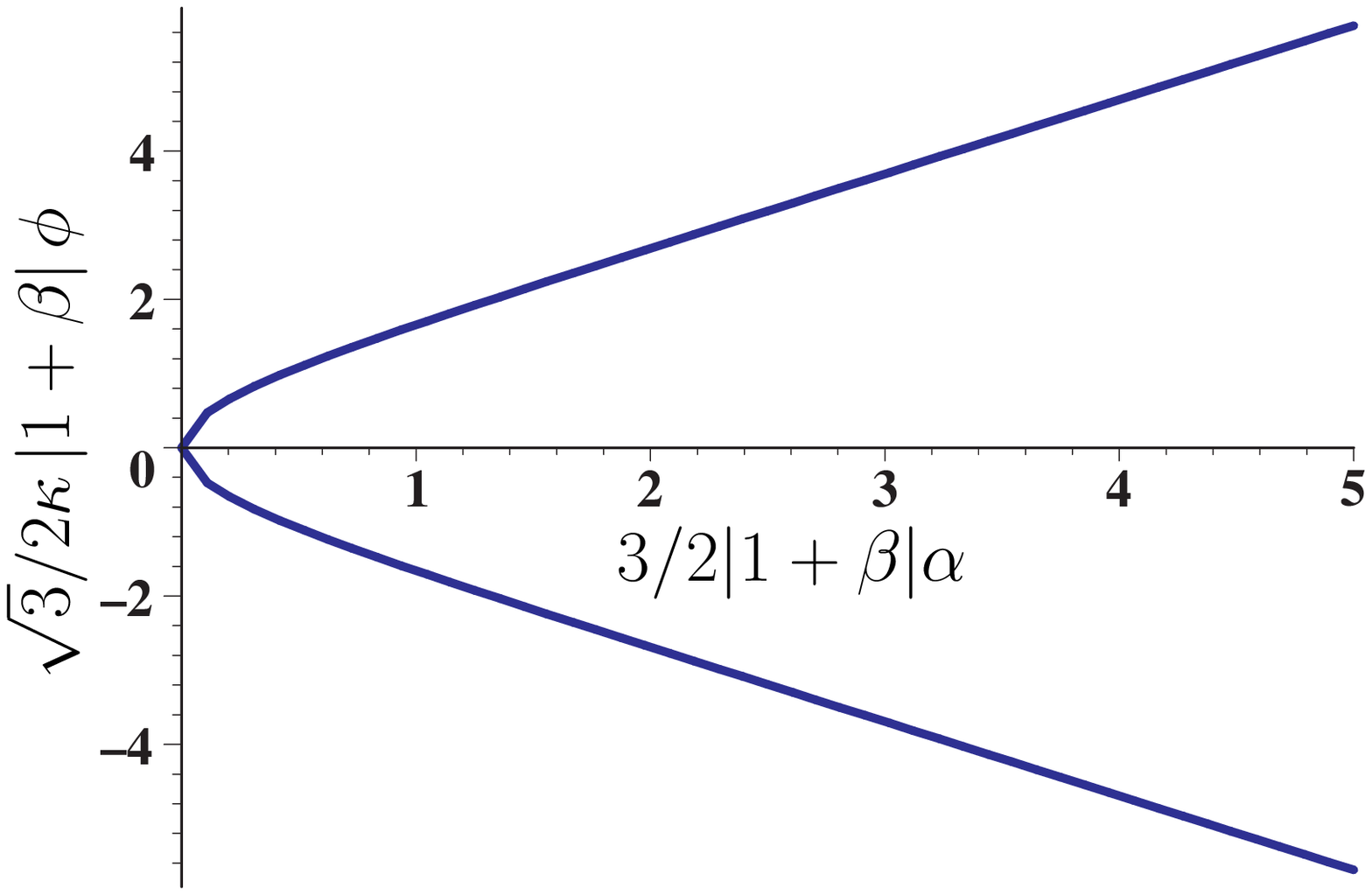}
\end{center}
\caption{\label{fieldversusscale-2a}Plot of the scalar field versus
the logarithmic scale factor $\alpha=\ln({a}/{a_{0}})$, where
$a_0$ corresponds to the location of the singularity in the
respective models. This  plot corresponds to a past big freeze
singularity where $a_0=a_{\rm min}$ (see Eq.~(\ref{phia})).}
\end{figure}

We first show how the GCG given in Subsection \ref{bfnonphantom} can
be described by a standard minimally coupled scalar field whose energy
density and pressure in a homogeneous and isotropic universe read
\begin{equation}
  \rho_\phi= \frac12 \dot\phi^2 + V(\phi)\ , \quad
  p_\phi=\frac12 \dot\phi^2 - V(\phi)\ .
\end{equation}
The dot corresponds
 to the derivative with respect to cosmic time. Then by imposing that
$\rho_\phi$ and $p_\phi$ satisfy the equation of state
(\ref{eqstate}), the kinetical energy density and the scalar field
potential evolve with the scale factor as
\begin{equation}
  \dot\phi^2=|A|^{\frac{1}{1+\beta}}\frac{\left(\frac{a_{\rm{min}}}{a}\right)^{3(1+\beta)}}
  {\left[\left(\frac{a_{\rm{min}}}{a}\right)^{3(1+\beta)}-1\right]^{\frac{\beta}{1+\beta}}}
  , \qquad
  V(a)=\frac12|A|^{\frac{1}{1+\beta}}\frac{\left(\frac{a_{\rm{min}}}{a}\right)^{3(1+\beta)}-2}
  {\left[\left(\frac{a_{\rm{min}}}{a}\right)^{3(1+\beta)}-1\right]^{\frac{\beta}{1+\beta}}}\
  .
  \label{keva}
\end{equation}
Therefore, the scalar field changes with the scale factor as (cf.
Figure~\ref{fieldversusscale-2a})

\begin{equation}
  |\phi-\phi_{\rm{min}}|(a)=\frac{2\sqrt{3}}{3\kappa|1+\beta|}\ln\left[\left(\frac{a_{\rm{min}}}{a}\right)^{\frac32\,(1+\beta)}+\sqrt{\left(\frac{a_{\rm{min}}}{a}\right)^{3(1+\beta)}-1}\,\right]\
  ,
  \label{phia}
\end{equation}
where $\phi_{\rm{min}}$ corresponds to the value of the scalar field
at $a_{\rm{min}}$, where the singularity occurs; we shall set
$\phi_{\rm min}=0$ for simplicity. Finally, by
combining (\ref{keva}) and (\ref{phia}), the scalar field potential reads

\begin{equation}
  V(\phi)=V_1\left[\sinh^{\frac{2}{1+\beta}}\left(\frac{\sqrt{3}}{2}\kappa|1+\beta||\phi|\right)
    -\frac{1}{\sinh^{\frac{2\beta}{1+\beta}}\left(\frac{\sqrt{3}}{2}\kappa|1+\beta||\phi|\right)}
  \right],
  \label{vphi}
\end{equation}
where $V_1=|A|^{\frac{1}{1+\beta}}/2$.
Notice that at the minimum scale factor or at $\phi=0$ the potential
is negative and divergent. In fact, the potential can be approximated
in this region by\footnote{At the lowest order, the first
term in (\ref{vphi}) does not contribute to the
approximation made in (\ref{vphismall}) because $-2\beta/(1+\beta)<2/(1+\beta)<0$.}

\begin{equation}
  V(\phi)\simeq -V_1\left(\frac{\sqrt{3}}{2}\kappa|1+\beta||\phi|\right)^{-\frac{2\beta}{1+\beta}}.
  \label{vphismall}
\end{equation}
We recall that $1+\beta<0$ and therefore $-2\beta/(1+\beta)< -2$;
that is, $V(\phi)$ is at the big-freeze singularity
more singular than an inverse square potential. This is  crucial for
the discussion of the quantum theory below.
On the other hand, for large values of the
scale factor or at large values of the scalar field, we have
\begin{equation}
\label{vphilarge}
  V(\phi)\simeq {2^{-\frac{2}{1+\beta}}} V_1\exp\left(-\sqrt{3}\kappa|\phi|\right).
\end{equation}
\begin{figure}[t]
\begin{center}
\includegraphics[width=8cm]{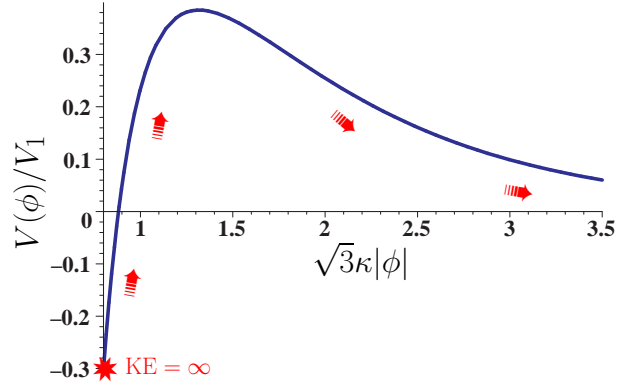}
\end{center}
\caption{Plot of the potential (\ref{vphi})
as a function of the scalar field for the value
    $\beta=-3$. The star denotes the location of the singularity,
    $\rm{KE}$ denotes the kinetic energy of the scalar field
    and the arrows denote the trajectory of the scalar field. }
\label{potentialplot-3a}
\end{figure}

The general behavior of the potential is shown in
Figure~\ref{potentialplot-3a}. The scalar field starts
with an infinite kinetic energy (at the past big-freeze singularity)
climbing up the potential until it reaches its top
and then starts rolling down the potential hill. When the
potential vanishes, the scale factor is equal to $a_{\rm{dom}}$.
Hence, the dominant energy density is violated when the scalar field
takes a value such that the potential becomes negative.

\subsection{The big-freeze singularity driven by a phantom scalar
field}
\label{bfphantomscalar}

Similarly to the case just considered,
the GCG discussed in Subsection \ref{bfphantom} can also be
described by a minimally coupled scalar field, although in this case
the phantom nature of the GCG implies that the scalar field does not have
the standard kinetic term; therefore, its energy density and pressure
read
\begin{equation}
  \rho_\phi= -\frac12 \dot\phi^2 + V(\phi)\ , \quad
  p_\phi=-\frac12 \dot\phi^2 - V(\phi)
\end{equation}
for a FLRW universe. Equating the previous quantities to
(\ref{rhobfp}) and (\ref{pbfp}), we obtain

\begin{equation}
  \dot\phi^2=A^{\frac{1}{1+\beta}}\frac{\left(\frac{a_{\rm{max}}}{a}\right)^{3(1+\beta)}}
  {\left[1-\left(\frac{a_{\rm{max}}}{a}\right)^{3(1+\beta)}\right]^{\frac{\beta}{1+\beta}}}
  , \qquad
  V(a)=\frac12A^{\frac{1}{1+\beta}}\frac{2-\left(\frac{a_{\rm{max}}}{a}\right)^{3(1+\beta)}}
  {\left[1-\left(\frac{a_{\rm{max}}}{a}\right)^{3(1+\beta)}\right]^{\frac{\beta}{1+\beta}}}\
  .
  \label{keva2}
\end{equation}
Consequently,
\begin{equation}
  \phi=\pm\frac{2}{\kappa\sqrt{3}}\frac{1}{1+\beta}\arccos\left[\left(\frac{a_{\rm{max}}}{a}\right)^{\frac{3(1+\beta)}{2}}\right],
  \label{phia2}
\end{equation}


\begin{figure}[t]
\begin{center}
\includegraphics[width=8cm]{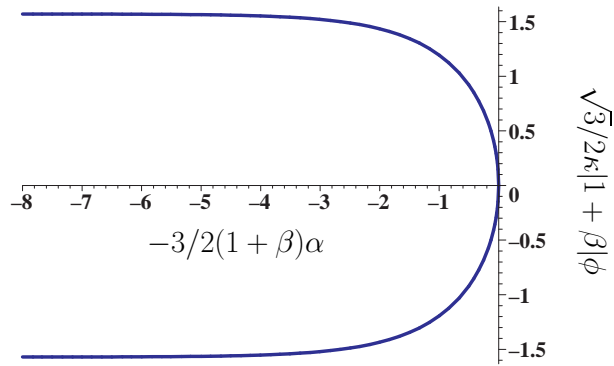}
\end{center}
\caption{\label{fieldversusscale-2b}Plot of the scalar fields versus
the logarithmic scale factor $\alpha=\ln({a}/{a_{0}})$, where
$a_0$ corresponds to the location of the singularity in the
respective models.  The  plot corresponds to a future big freeze/(past)big
\textit{d\'emarrage} singularity (see Eq.~(\ref{phia2})). In the
scenario with  a future big freeze or a
(past) big \textit{d\'emarrage} singularity, the quantity
$-3/2(1+\beta)\alpha$ is always  negative.}
\end{figure}

\begin{figure}[t]
\begin{center}
\includegraphics[width=8cm]{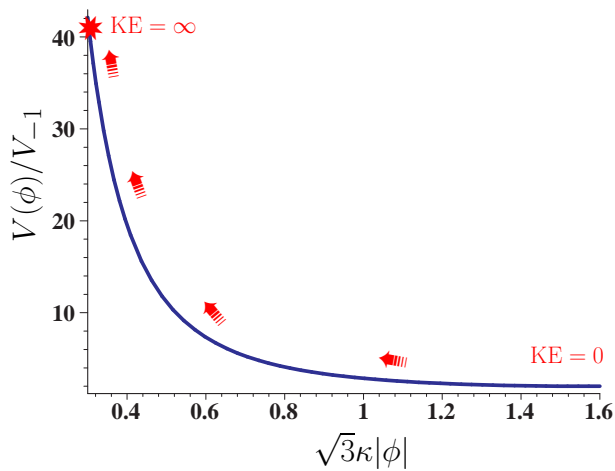}
\end{center}
\caption{Plot of the potential defined in
    (\ref{vphi2}) as a function of the scalar field for the value
    $\beta=-3$. The star denotes the location of the singularity,
    $\rm{KE}$ denotes the kinetic energy of the scalar field
    and the arrows indicate the trajectory of the scalar field.}
\label{potentialplot-3b}
\end{figure}

 In Figure~\ref{fieldversusscale-2b} we show how the scalar field is
 correlated
 with the scale factor. From the last equation one can read off that
 the scalar field vanishes when the
scale factor reaches its maximal classically allowed value. By
using the relations (\ref{keva2}) and (\ref{phia2}) we find
the following expression for the potential:
\begin{equation}
  V(\phi)=V_{-1}\left[\frac{1}{\sin^{\frac{2\beta}{1+\beta}}\left(\frac{\sqrt{3}}{2}\kappa|1+\beta||\phi|\right)}+
    \sin^{\frac{2}{1+\beta}}\left(\frac{\sqrt{3}}{2}\kappa|1+\beta||\phi|\right)
  \right],
  \label{vphi2}
\end{equation}
where $V_{-1}=A^{\frac{1}{1+\beta}}/2$ and
$0<(\sqrt{3}/{2})\kappa|1+\beta||\phi|\leq \pi/2$ (see
Figure~\ref{potentialplot-3b}).

The big-freeze singularity at $a=a_{\rm{max}}$ is now
located at $\phi=0$ where the scalar field potential can be
approximated by\footnote{Again, the second term in (\ref{vphi2})
does not contribute to the potential (\ref{vphismall2}) because
$-2\beta/(1+\beta)<2/(1+\beta)<0$.}
\begin{equation}
  V(\phi)\simeq V_{-1}\left(\frac{\sqrt{3}}{2}\kappa|1+\beta||\phi|\right)^{-\frac{2\beta}{1+\beta}}.
  \label{vphismall2}
\end{equation}
This potential (also the exact potential (\ref{vphi2})) can be deduced
from the expression (\ref{vphismall}) (resp. (\ref{vphi})) by rotating
$\phi\rightarrow i\phi$ and taking into account that $A$ changes
sign. Notice that this implies that there is a change of sign of the
potential (\ref{vphismall}) after performing such an analytical
continuation.

In general, the potential (\ref{vphismall2}) corresponds to a singular
potential. However, for large enough $\beta$ (and again for
$\phi\rightarrow 0$), $V(\phi)$ behaves effectively as an inverse
square potential; that is,
\begin{equation}
  V(\phi)\simeq
  V_{-1}\left(\frac{\sqrt{3}}{2}\kappa|1+\beta||\phi|\right)^{-2},\quad
  1\ll|\beta|.
\end{equation}
The scalar field starts its cosmological evolution with a vanishing
kinetic energy density at
$\sqrt{3}/{2}\kappa|1+\beta||\phi|=\pi/2$, then it climbs up through
the potential reaching an infinite kinetic energy when the classical
universe reaches its maximum size, that is, when $\phi\rightarrow 0$
[see Figure~\ref{potentialplot-3b}]. This is a usual feature
of phantom scalar fields. Notice as well that the standard canonical
scalar field with the potential (\ref{vphi}) also climbs through its
potential but in this case the reason behind this strange behavior
is the initial infinite kinetic energy of the scalar field when
$\phi\rightarrow 0$.

\subsection{The  big-\textit{d\'emarrage} singularity driven by a
  phantom scalar field}\label{bigdemarrage-sf}

\begin{figure}[t]
\begin{center}
\includegraphics[width=8cm]{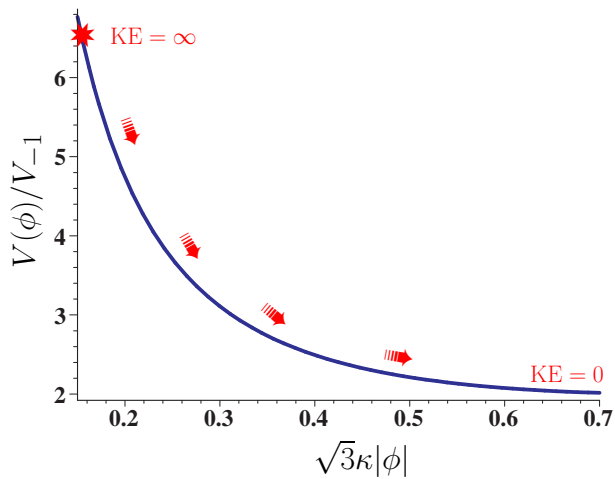}

\end{center}
\caption{Plot of the potential defined in
    (\ref{vphi2}) as a function of the scalar field for the value
    $\beta=3$. The star denotes
     the location of the singularity, $\rm{KE}$ denotes the kinetic
     energy of the scalar field and the arrows indicate the trajectory of the scalar field. }
\label{potentialplot-3c}
\end{figure}

Similar to the above cases one can also show that the phantom GCG
leading to a big-{\em d\'emarrage} singularity can be mimicked by a
minimally coupled phantom scalar field $\phi$, where $\phi$ and the
scalar field potential $V(\phi)$ are given, respectively, by
Eqs~(\ref{phia2}) and (\ref{vphi2}) after substituting
$a_{\rm{max}}$ by $a_{\rm{min}}$, see also
Figures~\ref{fieldversusscale-2b} and \ref{potentialplot-3c}. The
expressions given in (\ref{keva2}) are also valid after exchanging
$a_{\rm{max}}$ by $a_{\rm{min}}$. The scalar field starts moving
away from the singularity (located at $\phi=0$ or $a=a_{\rm{min}}$)
by rolling down the potential with an infinite kinetic energy. The
phantom nature of the scalar field implies that $\phi$ loses its
kinetic energy as it rolls down the potential, cf.
Figure~\ref{potentialplot-3c}. In fact, when the scalar field
reaches the tail of the potential, that is,
$\sqrt{3}/{2}\kappa|1+\beta||\phi|\rightarrow\pi/2$, its kinetic
energy is approaching zero, which is in agreement with the
asymptotic de Sitter-type behavior of a universe filled by this
phantom GCG.

Close to the singularity ($\phi\rightarrow 0$), $V(\phi)$ can be
approximated by (\ref{vphismall2}). Because $\beta>0$, this
potential does not correspond to a singular potential, although for
$1\ll \beta$ it behaves as an inverse square potential. Notice as well
that for $\beta=1$ the potential behaves as a Coulomb potential like
its non-phantom version \cite{Kamenshchik:2007zj}.


\section{Wheeler--DeWitt equation and quantum states}
\label{QC}

In this section, we shall describe the quantization  of the classical
scenarios discussed above. This  will be carried out in the quantum
geometrodynamical framework. Its central equation is the
Wheeler--DeWitt equation, depending on the configuration space variables
$(a,\phi)$. From the solutions obtained, we shall retrieve
information concerning  the above GCG models with regard
to their quantum behavior at the classical singularity\footnote{The
  generalized Chaplygin has been 
  previously analyzed from a quantum point of view in different setups 
  (see Ref.~\cite{BouhmadiLopez:2004mp}).}. A
discussion of the influence of different boundary conditions on the
wave function is presented in the next section.

Before proceeding to more technical aspects, let us mention that
even though singularity avoidance is a major touchstone for any
quantum gravity theory, no consensus regarding the criteria that may
account for such an avoidance exists. This is mainly due to the fact
that the criteria that one admits depend strongly on how one
interprets the  wave function of the universe. Divergent opinions on
that interpretation arise because quantum gravity is itself a
timeless theory \cite{OUP}. In this respect it differs from any
other quantum field theory, and there is not yet any agreement about
the appropriate interpretational framework.

One possible line to establish a reasonable research criterium is to
 regard the quantum-cosmological wave function as the fundamental
entity from which our spacetime can be derived in an appropriate
limit. The recovery of spacetime can be expected to occur only in
special regions of configuration space. The derivation of the
semiclassical limit is performed by a Born--Oppenheimer type of
approximation scheme with decoherence as an essential ingredient,
cf. \cite{OUP} and the references therein. The gravitational part of
the wave function is usually taken to be of a WKB form, which means
that narrow wave packets around classical trajectories would not
spread. But wave packets which are initially peaked around classical
trajectories may not remain so along the entire trajectory. Such a
dispersion is unavoidable in a quantum universe whose classical
version exhibits recollapse \cite{Kiefer1988}. Thus, a spreading of
the wave packet signals a breakdown of the semiclassical
approximation; one can then no longer associate with the wave
function a classical spacetime as an approximate concept. Such a
spreading necessarily occurs when approaching the region of the
classical big-rip singularity in quantum phantom cosmology, a
phenomenon which we interpreted in \cite{DKS} as singularity
avoidance.

Another sufficient (but by no means necessary) criterium for
singularity avoidance is the vanishing of the wave function at the
classical singularity; there is then no possibility for such a
singularity to occur in any limit. Vanishing of the wave function as
a criterium for singularity avoidance was first suggested by Bryce
DeWitt in his pioneering paper \cite{DeWitt1967}. Singularity
avoidance in this sense occurred for the big-brake singularity
\cite{Kamenshchik:2007zj}. Vanishing of the wave function will also
be in the present section the appropriate criterium for singularity
avoidance.

Let us, then, turn to a detailed analysis of the quantum versions for
the above discussed classical models.
The wave function satisfies the Wheeler--DeWitt
equation, which with the Laplace--Beltrami factor ordering reads

\begin{equation}
  \frac{\hbar^2}{2}\left(\frac{\kappa^2}{6}\frac{\partial^2}{\partial\alpha^2}-\ell\frac{\partial^2}{\partial\phi^2}\right)
  \Psi\left(\alpha,\phi\right)+
  a_0^6e^{6\alpha}V(\phi)\Psi\left(\alpha , \phi\right)\nonumber =0,
  \label{WdW1}
\end{equation}
where $V(\phi)$ is given in (\ref{vphi}) for the standard scalar
field model and in (\ref{vphi2}) for both phantom scalar field
models. We have introduced the new variable
$\alpha:=\ln\left(\frac{a}{a_0}\right)$ and assume that $a_0$
corresponds to the location of the singularity in the respective
models. In the following, we shall use $\tilde{a}:=\frac{a}{a_0}$
instead of $a$ such that $\tilde{a}_0=1$ holds. For
simplicity, we
shall drop the tilde. Recall that $\kappa^2=8\pi G$. We have
introduced the parameter $\ell$ in order to distinguish between the
phantom and non-phantom scalar field. For the phantom scalar field we have
$\ell=-1$, whereas we have $\ell=1$ for ordinary scalar-field matter. Note
that a fundamental length scale is necessary in the Wheeler--DeWitt
equation to yield correct dimensions. To solve this
equation, we make the \emph{ansatz}
\be \Psi(\alpha,\phi)=\varphi_k(\alpha,\phi)C_k(\alpha)\ , \ee
where $k$ is a priori not restricted to real values.  Furthermore,
we require $\varphi_k$ to satisfy
\be
-\ell\frac{\hbar^2}{2}\frac{\del^2\varphi_k}{\del\phi^2}+a_0^6e^{6\alpha}V(\phi)\varphi_k=E_k(\alpha)\varphi_k\
.  \ee 
Such a Born--Oppenheimer-type of ansatz was first used in quantum
cosmology in \cite{Kiefer1988} in the study of wave packets. It
assumes the approximate validity of a quasi-separability between
$\alpha$ and $\phi$, that is, the matter part $\varphi_k$ depends only
adiabatically on the scale factor. We discuss the validity of the
Born--Oppenheimer approximation in Appendix~A.

As the potentials (\ref{vphi}) and (\ref{vphi2}) are rather complicated,
we solve the Wheeler--DeWitt equation for certain ranges of
$\vert\phi\vert$ and approximate the respective potential there. For
the study of singularity avoidance the region of primary interest is
$\vert\phi\vert\ll 1$. This corresponds in each model
to the vicinity of the singularity. In this region, the potential
is approximated
by (\ref{vphismall}) for the ordinary scalar field and by
(\ref{vphismall2}) in the case of the phantom field.
Introducing the notation
\be
V_{\alpha}:=a_0^6e^{6\alpha}V_\ell\left[\frac{\sqrt{3}\kappa}{2}\vert1+\beta\vert\right]^{-\frac{2\beta}{1+\beta}}\
, \ee
we find that $\varphi_k$ has to satisfy
\be -\ell\frac{\hbar^2}{2}\frac{\del^2\varphi_k}{\del\phi^2}-\ell
V_{\alpha}\vert\phi\vert^{-\frac{2\beta}{1+\beta}}\varphi_k=E_k(\alpha)\varphi_k\
.  \ee
Defining $k^2:=\frac{2E_k}{\hbar^2}$
and $\tilde{V}_{\alpha}:=\frac{2V_\alpha}{\hbar^2}$, we finally arrive at
\be \lb{varphieqn} \varphi_k^{\prime \prime}+\left[\ell
 k^2+\tilde{V}_{\alpha}\vert\phi\vert^{-\frac{2\beta}{1+\beta}}\right]\varphi_k=0\
, \ee
where $^\prime$ denotes a
derivative\footnote{In fact, there is one
such equation for positive and
  one for negative $\phi$. Both equations are (due to the
  modulus-dependence of the potential) identical. Matching is
  carried out through the conditions $\Psi_+(0)=\Psi_-(0)$ and
  $\Psi^\prime_+(0)=-\Psi^\prime_-(0)$ where $\Psi_\pm$ refer to the
  positive/negative $\phi$-solutions, respectively. These conditions
  imply that the constants in both solutions coincide. We will
  therefore just refer to the modulus-dependent equation (\ref{varphieqn})
  and its solution, keeping in mind that we have the  same solution
  for positive and negative $\phi$.} with respect to $\phi$. We
recognize that this equation is formally the same as the radial part
of the stationary Schr\"odinger equation for an {\em attractive}
potential of inverse power $V\sim r^{-\frac{2\beta}{1+\beta}}$,
where $\vert\phi\vert$ plays the role of the radial coordinate $r$,
and the angular momentum vanishes.

Equation (\ref{varphieqn}) is the central equation for the following
discussion. Because it formally resembles the Schr\"odinger equation,
we are able to make use of results encountered in quantum
mechanics. Potentials of the type $V\sim
r^{-\frac{2\beta}{1+\beta}}$ are there called {\em singular}
\cite{Frank:1971xx,Case:1950an}. Any potential that approaches
(plus or minus) infinity faster than $r^{-2}$ for $r\to 0$ belongs to
this class. For an attractive $r^{-2}$-potential there exists a
transitional case: if the coupling is more negative than a critical
value, the potential is singular, otherwise regular.

Analytical solutions for polynomial singular potentials are known for
the inverse
square, inverse fourth-power, and inverse sixth-power potentials. The
inverse square potential is realized for $\vert\beta\vert\gg 1$,
where $\beta$ is chosen such that $\vert1+\beta\vert\vert\phi\vert$
is still small. The inverse fourth-power potential corresponds to
$\beta=-2$, whereas the inverse sixth-power potential is realized for
$\beta=-\frac 32$.

In our paper we shall focus on the case
$\vert\beta\vert\gg 1$ for the standard scalar field as well as the
two phantom field models. We thus deal with the case of the inverse-square
potential $\frac{\tilde{V}_{\alpha}}{{\vert\phi\vert}^2}$ with
\bdm
\tilde{V}_{\alpha}=\frac{2a_0^6 e^{6\alpha}V_{\ell}}{\hbar^2}
\left[\frac{\sqrt{3}\kappa\vert\beta\vert}{2}\right]^{-2}>0 \ .
\edm
The $r^{-2}$-potential in quantum mechanics was discussed, for
example, in \cite{Frank:1971xx,Case:1950an,Narnhofer}, cf. also
\cite{polymer} and the references therein. It became recently of
interest in studying the polymer quantization which is motivated by
loop quantum gravity \cite{polymer}.

Note that for the models with big-freeze
singularity we then have $\beta\ll-1$, whereas for the model with
big-d\'emarrage singularity we have $\beta\gg1$. We expect that
this case is sufficiently generic to accommodate also the features of
other singular potentials. As is known from quantum mechanics
\cite{Frank:1971xx,Case:1950an}, the most important issue for singular
potentials is the fact that square integrability does no longer
suffice to select a unique class of states and that the spectrum
therefore contains an ambiguity. This ambiguity points
to new physics at short scales and must be fixed by either experiment
or knowledge from a more fundamental theory. It is the fortunate
property of the Coulomb potential that its ensuing states do {\em not}
contain such an ambiguity (see, however, \cite{Fewster}).
Thus, for the least singular potential, which is realized for
$\vert\beta\vert\gg1$, we have to solve the equation
\be \varphi_k^{\prime \prime}+\left[\ell
 k^2+\frac{\tilde{V}_{\alpha}}{{\vert\phi\vert}^2}\right]\varphi_k=0\
\label{30}
.  \ee
Note that phantom and scalar matter have to obey the same quantum
equation\footnote{Note that $V_1=\frac{\vert
    A\vert{}^{\frac{1}{1+\beta}}}{2}$ and
  $V_{-1}=\frac{A^{\frac{1}{1+\beta}}}{2}$, so both are positive and coincide
  numerically.}, where the realm of positive energy for the ordinary
scalar field $k^2>0$ corresponds to the realm of negative energy for
the case of the phantom field, $k^2<0$, cf. Eq.~(\ref{varphieqn}). The
general solution to this equation is given by \cite{Abramowitz}
\be\label{generalvarphik}
\varphi_k(\alpha,\vert\phi\vert)=\sqrt{\vert\phi\vert}\left[c_1\mathrm{J}_\nu(\sqrt{\ell}k\vert\phi\vert)+c_2\mathrm{Y}_\nu(\sqrt{\ell}k\vert\phi\vert)\right]\
, \ee where $\nu:=\sqrt{\frac14-\tilde{V}_\alpha}$, so the index is
a function of $\alpha$. There are four cases to distinguish: $k$ can
be real or imaginary, depending on whether the energy entering $k^2$
is positive or negative. Furthermore, $\nu$ can be real or
imaginary, depending on the parameters $\beta$,  $A$, and the value
of $\alpha$. Note that $\varphi_k$ satisfies
\be
\label{orthogonality}
\ell(k^2-n^2)\int_0^b{\rm d}
\phi\ \varphi_k(\phi)\varphi_n(\phi)=\left[\varphi_k(b)\frac{{\rm d}
    \varphi_n}{{\rm d}\phi}(b)-\varphi_n(b)\frac{{\rm d} \varphi_k}
{{\rm d}\phi}(b)\right]\ ,
\ee
where $b=\infty$ for the standard scalar field and
$b=\phi_\star:=\frac{\pi}{2}\frac{2}{\sqrt3\kappa\vert1+\beta\vert}$
for the phantom
scalar field. (This equation is analogous to Equation (11) in
\cite{Case:1950an}.) We have fixed the range of definition for $\phi$
in the phantom case from $0$ to $\phi_\star$ because all classical
solutions are restricted to this range. That this also holds in the
quantum theory is, of course, an assumption, similar to treating, say,
a particle on the half-line only.

The lower bound in the integral (\ref{orthogonality}) does not
contribute as $\varphi_k$
vanishes at the origin (see below). Thus the matter dependent part of
the wave function is
not necessarily orthogonal. But it is if the right-hand side of
(\ref{orthogonality}) vanishes. This is the case if
$b=\infty$ and $\nu$ is not an integer, as happens for the standard scalar field. The reason is
the following: (i) The first
Bessel function ${\rm J}_{\nu}$ satisfies an orthogonality
relation and (ii) the $\mathrm{J}_\nu(\sqrt{\ell}k\vert\phi\vert)$ and
$\mathrm{J}_{-\nu}(\sqrt{\ell}k\vert\phi\vert)$
are linearly independent solutions of (\ref{30}) (except when $\nu$ is
an integer). Therefore, as  the solution $\varphi_k$ can then be expressed
exclusively in terms of the first kind of Bessel functions,  $\varphi_k$
is orthogonal if either of the two coefficients in front of
$\mathrm{J}_\nu(\sqrt{\ell}k\vert\phi\vert)$ or
$\mathrm{J}_{-\nu}(\sqrt{\ell}k\vert\phi\vert)$ is zero.
Otherwise, that is, if $\nu$ is an integer,
additional conditions have to be imposed to get
orthogonal eigenfunctions,
as it is the case for the phantom scalar field.

Different types of solutions for the three models discussed in our paper will
differ by the boundary conditions that we may choose to impose (see
next section). This choice is determined by the classical
trajectory. For ordinary matter, the classical trajectory has a
minimum (see Figure~2),
whereas for phantom matter with $\beta+1<0$ it reaches its
maximum at the classical singularity (see Figure~3).
 For the second phantom model
with $\beta>0$, the singularity lies again at a minimum of the
classical trajectory. We will see how this difference in the
classical model influences the quantum behavior through the
boundary condition. Moreover, the phantom field is restricted to a
finite range.

Independent of boundary conditions,
the singularity occurs in all three models at $\phi=0$
and $\alpha=0$. For $\alpha=0$,
$\nu=\nu_0:=\sqrt{\frac14-\tilde{V}_{\alpha=0}}$. There are three
cases to distinguish, $\frac14-\tilde{V}_{\alpha=0}>0$,
$\frac14-\tilde{V}_{\alpha=0}<0$ and the transitional case
$\frac14-\tilde{V}_{\alpha=0}=0$ . In the first and third case, $\nu_0$ is
real, whereas in the second case it is purely imaginary.
\begin{itemize}
\item$\frac14-\tilde{V}_{\alpha=0}>0$.\\ Then $0<\nu<\frac12$.
In the quantum mechanical analogy, this corresponds to the case of a
regular potential. We get
from the behavior of the first Bessel function for small argument,
  \be
 \mathrm{J}_\nu(z)\approx\left(\frac{z}{2}\right)^\nu\frac{1}{\Gamma(\nu+1)},\
  \qquad \nu\neq -1,-2,-3\ldots,\label{approxJnu}
 \ee cf. \cite{Abramowitz}, Eq.~9.1.7, that it vanishes at the origin.
This holds for real as well as imaginary argument.
The behavior of the second Bessel function for small argument is
given by
\be
 \mathrm{Y}_\nu(z)\approx-\frac1\pi\left(\frac{z}{2}\right)^{-\nu}\Gamma(\nu),\
  \qquad \rm{Re}(\nu)>0,\label{approxYnu}
\ee cf. \cite{Abramowitz}, Eq.~9.1.9. Thus it diverges weaker than
$\vert\phi\vert^{-\frac12}$. As the Bessel functions in $\varphi_k$
are multiplied by a factor $\sqrt{\vert\phi\vert}$, this divergence
is, however,
cancelled. We thus conclude that the matter-dependent part of the wave function
vanishes at the singularity.

In the quantum mechanical case of regular potentials, one of the two
independent solutions of the Schr\"odinger equation is not
normalizable and therefore discarded. This is, however, not
necessarily the case for vanishing angular momentum, since for the
Coulomb potential, for example, both solutions are normalizable. The
selection occurs in that case because only one of the solutions leads
to an essential self-adjoint Hamiltonian \cite{Behncke}. Such an
argument can, however, not be invoked in the present case because
the classical time parameter $t$ is absent in quantum cosmology. There
is thus no notion of unitarity here and thus no reason to advocate the
self-adjointness of the Hamiltonian. We thus have non-uniqueness also
in the ``regular'' case.

\item$\frac14-\tilde{V}_{\alpha=0}<0$.\\ This corresponds to the
  singular case in quantum mechanics.
Here, $\nu_0$ is
  imaginary; we write $\nu_0=\mathrm{i}\nu_0$. Since the above formula
  (\ref{approxJnu}) holds for $\nu\neq -1,-2,-3\ldots$,
we can still use it and find
\be
\lim_{\vert\phi\vert\to0}\sqrt{\vert\phi\vert}\mathrm{J}_{\mathrm{i}\nu_0}(\sqrt{\ell}\vert\phi\vert
  k)=\lim_{\vert\phi\vert\to0}\sqrt{\vert\phi\vert}\frac{1}{\Gamma(1+\mathrm{i}\nu_0)}e^{\mathrm{i}\nu_0\ln\left(\frac{\sqrt{\ell}k\vert\phi\vert}{2}\right)}\ .
\ee
For real as well as imaginary $k$ (corresponding to positive and
negative energy, respectively), this part of the matter-dependent wave function
oscillates infinitely rapidly as it goes to zero as
$\vert\phi\vert\to0$.
For the second part of the matter-dependent wave function we
introduce a small parameter $\epsilon>0$, since
(\ref{approxYnu}) holds only for $\rm{Re}(\nu)>0$,
\be
 \lim_{\vert\phi\vert\to0}\lim_{\epsilon\to0}\sqrt{\vert\phi\vert}\mathrm{Y}_{\epsilon+\mathrm{i}\nu}(\vert\phi\vert
 \sqrt{\ell}
 k)=-\frac1\pi\lim_{\vert\phi\vert\to0}\lim_{\epsilon\to0}\sqrt{\vert\phi\vert}\left(\frac{\sqrt{\ell}k\vert\phi\vert}{2}\right)^{-\epsilon}\Gamma(\epsilon+\mathrm{i}\nu_0)e^{\mathrm{i}\nu_0\ln\left(\frac{\sqrt{\ell}k\vert\phi\vert}{2}\right)}\
 .
  \ee
This function oscillates very rapidly but goes to zero as
$\vert\phi\vert\to0$ due to the factor $\sqrt{\vert\phi\vert}$.
We therefore conclude that the wave function vanishes at the origin in
this parameter range as well.
\item$\frac14-\tilde{V}_{\alpha=0}=0$.\\ In this case we have to  consider the Bessel functions $\mathrm{J}_0(z)$ and
  $\mathrm{Y}_0(z)$ for small $z$. It is $\mathrm{J}_0(0)=1$ and so
  this part of $\varphi_k$ vanishes when multiplied by
  $\sqrt{\vert\phi\vert}$for $\phi\to0$. But
  $\mathrm{Y}_0(z)\approx\frac2\pi\ln{\left(z\right)}$ for small $z$, \cite{Abramowitz}, 9.1.8, and thus
  diverges but
  $\sqrt{z}\mathrm{Y}_0(z)\approx\frac2\pi\sqrt{z}\ln(z)\to 0$
 as $z\to0$. We thus find a vanishing wave function in this case as well.
\end{itemize}

The singularities in these three cases are thus avoided when  the
matter-dependent part of the wave function vanishes. To complete
this claim we have to insure that the
gravitational part of the wave function does not diverge at the
respective singularities. Note that this result does not depend on
any boundary conditions but is just a consequence of the
Wheeler--DeWitt equation.

Note also that the general solution $\varphi_k$
does not vanish at the singularity if $\nu=\frac12$. 
This corresponds to $V_{\ell}=0$, thus a vanishing potential. This case is
excluded by the set-up of our model. One could read this as: no
potential implies no singularity and thus no singularity avoidance but
a finite wave function at the origin $\phi=0$. Nevertheless, it is
always possible to pick up a specific solution, that is, a particular
$\varphi_k$ which vanishes.

Concerning the gravitational part of the wave function, inserting
the solution (\ref{generalvarphik}) for the matter-dependent part of
the wave function into the Wheeler--DeWitt equation (\ref{WdW1}), we
arrive at
\be
\label{WDWvarphi}
\frac{\kappa^2}{6}\left(2\dot
    C_k\dot\varphi_k+C_k\ddot\varphi_k\right)+\left(\frac{\kappa^2}{6}\ddot
    C_k+ k^2C_k\right)\varphi_k=0\ , \ee
where a dot denotes derivative with respect to $\alpha$. To obtain
the gravitational part of the wave function, we assume
that the terms $\dot
C_k\dot\varphi_k$ and $C_k\ddot\varphi_k$ can be neglected; this is
just the meaning of the Born--Oppenheimer approximation discussed above.
Basically, we assume that $C_k$ varies much more rapidly with
$\alpha$ than $\varphi_k$ and, moreover, neglect the back reaction
of the matter part on the
gravitational part. In summary, we are assuming  that the change in
the matter part does not influence the gravitational part; the
matter part simply contributes its energy through $k^2$.
This leads to the equation
\be\label{Ckeqn}
\left(\frac{\kappa^2}{6}\ddot
    C_k+ k^2C_k\right)\varphi_k=0\ .
\ee
It is solved by
\be\label{gravsolution}
C_k(\alpha)=b_1e^{\mathrm{i}\frac{\sqrt6k}{\kappa}\alpha}+b_2e^{-\mathrm{i}\frac{\sqrt6k}{\kappa}\alpha}\
.  \ee As the equation determining the gravitational part of the
wave function is independent of $\ell$, we get the same solution,
irrespective of whether we deal with a phantom or with a scalar
field. Note that for $k^2<0$, $k$ becomes imaginary and the
dependence on $\alpha$ becomes exponential. In any case,
$C_k(\alpha=0)<\infty$, so the wave function remains finite at the
respective singularities and we can safely speak of singularity
avoidance.


\section{Boundary conditions and singularity avoidance}\label{6}

Nobody knows what the correct boundary
  condition for the quantum universe are. There have been several
  proposals, most of them using the
  boundary condition with the ambition to lead to singularity avoidance
  \cite{DeWitt1967, Hartle:1983ai, Vilenkin, CZ91}. DeWitt, in particular,
  speculates that the fact that a boundary condition
  is generally needed to make the quantum theory `singularity-free' is
  an argument in favor of the theory: the theory itself does not
  leave any freedom of choice but provides the boundary condition
  itself \cite{DeWitt1967}.
As for the present state of understanding quantum gravity, however,
singularity avoidance is demanded from the outside as a selection
criterium.

Irrespective of this discussion, it is in general necessary to
impose a completely \emph{different} condition if one instead wants
to construct wave packets that follow classical trajectories with
turning point in configuration space \cite{Kiefer1988}. Namely, one
has to require that the wave packet decays in the classically
forbidden region. This allows the interference of wave packets
following the two branches of the classical solution behind the
classical turning point. Of course, this is just the standard
quantum mechanical treatment of classically forbidden regions. In
general, out of solutions to the Wheeler--DeWitt equation which grow
in the classically forbidden region, no wave packet can be
constructed that follows the classical path. This is what happens
generically for the no-boundary state \cite{CK91}. In order to make
connection with the underlying classical theory, we shall therefore
use herein as condition that the wave function decrease in the
classically forbidden region.

One cannot assume that the condition of the wave function to decrease
into the classically forbidden region always fixes it uniquely.
Nonetheless it is a necessary
condition that has to be imposed in order to get a theory with
correct semiclassical limit. It can be used in
 selecting physically sensible solutions out of the general set of
solutions to the Wheeler--DeWitt equation. We now want to explore the
situation for our models presented above.

\subsection{Defrosting the big freeze}

\subsubsection{Standard scalar field}\label{BCbfnonphantom}

In the classical scalar field model, the region
$a<a_{\rm{min}}=1$ is a forbidden region.\footnote{Recall that
we use $\tilde{a}=\frac{a}{a_{\rm{max}}}$ and drop the tilde.}
We impose the boundary
condition that the wave function decay there.
In terms of the variable $\alpha$, the boundary
condition to be imposed is therefore $\Psi\to0$ as
$\alpha\to-\infty$. But as $\alpha\to-\infty$, $\nu\to\frac12$. The
matter-dependent part of the wave function is then given by
\be
\lim_{\alpha\to-\infty}\varphi_k(\alpha,\vert\phi\vert)=\sqrt{\frac{2}{\pi}}\left[c_1\sin(k\vert\phi\vert)-c_2\cos(k\vert\phi\vert)\right]\ 
, \ee 
see \cite{Abramowitz}, Eqs.~10.1.11 and 10.1.12. This vanishes for small
$\vert\phi\vert$ if $c_2=0$ for both real and imaginary
$k$. Thus imposing  $c_2=0$, we are left with the solution\footnote{In
  the quantum mechanics of a repulsive $r^{-2}$-potential,
  $r^{1/2}{\rm J}_{\nu}(kr)$ is the unique solution which vanishes at
  $r=0$, cf. Equation (3.3) in \cite{Frank:1971xx}.}
\be \label{varphikscalar}
\varphi_k(\alpha,\vert\phi\vert)=c_1\sqrt{\vert\phi\vert}\mathrm{J}_\nu(k\vert\phi\vert)\
.  \ee Due to the vanishing of $\varphi_k$ at the origin and its boundedness
at infinity, the
$\varphi_k$ are orthogonal, that is,
 $\int{\rm d}\phi\varphi_k\varphi_n\propto\delta(k-n)$
 \cite{Arfken}. 
 This is due to the fact
that $\phi$ can become arbitrary large. If the integration range of
$\phi$ was finite, we would need additional conditions to ensure
orthogonality, cf. \cite{Abramowitz}, Eq.~11.4.5. Despite the
orthogonality, $k$ is not restricted to be of integer value.

Inserting the solution (\ref{varphikscalar}) for the
matter-dependent part of the wave function into the Wheeler--DeWitt
equation (\ref{WdW1}), we arrive at (\ref{WDWvarphi}). To obtain
the gravitational part of the wave function, we proceed as before
and find (\ref{gravsolution}) for the gravitational part
of the wave function.

For positive energy, $k^2>0$, the gravitational part of the wave
function is thus oscillating. No further complications arise and the
full solution is given by
\be
\Psi_k(\alpha,\phi)=c_1\sqrt{\vert\phi\vert}\mathrm{J}_\nu(k\vert\phi\vert)\left[b_1e^{\mathrm{i}\frac{\sqrt6k}{\kappa}\alpha}+b_2e^{-\mathrm{i}\frac{\sqrt6k}{\kappa}\alpha}\right]\
.\label{psik1}
\ee
For negative energy, however, the gravitational part becomes exponential. To ensure that the boundary condition $\Psi\to0$ as
$\alpha\to-\infty$ is satisfied for the entire wave function, we have
to set $b_1=0$. Thus, for imaginary $k$ the gravitational part of the wave
function decays exponentially for $\alpha\to-\infty$ whereas the
matter part remains finite, see Eq.~(\ref{varphikscalar}). The gravitational
part alone ensures in this way that the wave function vanishes as
$\alpha\to-\infty$. No additional condition arises for
$\varphi_k$. The full solution for imaginary $k\to \mathrm{i} k$ is thus
\be
\Psi_k(\alpha,\phi)=b_2e^{\frac{\sqrt6
    k}{\kappa}\alpha}\left[c_1\mathrm{J}_\nu(\mathrm{i}k\vert\phi\vert)+c_2\mathrm{Y}_\nu(\mathrm{i}k\vert\phi\vert)\right]\ ,\label{psik2}
\ee
and $\varphi_k$ and $\varphi_n$ are not orthogonal in this case.

Notice that the functions $\Psi_k(\alpha,\phi)$ given in (\ref{psik1}) and
(\ref{psik2}) fulfil as well the DeWitt criterium as
they vanish for $\alpha=0$ and $\phi=0$.

\subsubsection{Phantom scalar field}\label{BCbfphantom}

The classical model with the phantom-driven generalized Chaplygin gas
has a classically forbidden region given by $a>a_{\rm max}=1$. We
therefore impose the boundary condition $\Psi\to0$ as $\alpha\to\infty$.
In this limit, $\nu$ becomes purely imaginary and large.
We shall set here $\nu:=\mathrm{i}\nu$ and $\nu\to\infty$
($\nu$ now being real). We are thus looking for a combination of
Bessel functions which vanish for large imaginary index.

The wave function for positive energies, that is, $k^2>0$ can be written as
\begin{equation}
\varphi_k(\alpha,\vert\phi\vert)=\sqrt{\vert\phi\vert}\left[c_1\mathrm{J}_{\mathrm{i}\nu}(\mathrm{i}k\vert\phi\vert)+c_2\mathrm{Y}_ {\mathrm{i}\nu}(\mathrm{i}k\vert\phi\vert)\right]\ .
\end{equation}
As in this case the argument of the Bessel functions is purely imaginary, it is rather convenient to rewrite the wave function in terms of the linearly independent modified Bessel functions \cite{Abramowitz},
\begin{equation}
\varphi_k(\alpha,\vert\phi\vert)=\sqrt{\vert\phi\vert}\left[\tilde{c}_1\mathrm{I}_{\mathrm{i}\nu}(k\vert\phi\vert)+\tilde{c}_2\mathrm{K}_{\mathrm{i}\nu}(k\vert\phi\vert)\right]\ ,
\end{equation}
where $\tilde{c}_1,\tilde{c}_2$ are arbitrary constants. It can be shown that the modified Bessel function $\mathrm{K}_{\mathrm{i}\nu}(k\vert\phi\vert)$ goes asymptotically to zero as $\nu\to\infty$ by using
\begin{eqnarray}
\mathrm{K}_{\mathrm{i}\nu}&=&\sqrt{2}(\nu^2-x^2)^{-\frac14}
\exp\left(-\frac{\nu}{2}\pi\right)\times\left[\mathrm{const.}+\mathcal{O}((\nu^2-x^2)^{-\frac12})\right],\ \,\,\,  \nu>x>0,
\end{eqnarray}
cf. \cite{Bateman}, Eq.~ 7.13.2(19). Indeed, the modulus of the function $\mathrm{K}_{\mathrm{i}\nu}(k\vert\phi\vert)$ is oscillatory and its local extremum goes to zero as $\nu\to\infty$. Therefore, the implementation of the boundary condition gives
\be
\varphi_k(\phi)=\tilde{c}_2\sqrt{\vert\phi\vert}\mathrm{K}_{i\nu}(k\vert\phi\vert)\ . \label{positiveEphantom} \ee
On the other hand, the wave function for negative energies, that is,
$k^2<0$, can be written as
\begin{equation}
\varphi_k(\alpha,\vert\phi\vert)=\sqrt{\vert\phi\vert}\left[c_1\mathrm{J}_{\mathrm{i}\nu}(\tilde
  k\vert\phi\vert)+c_2\mathrm{Y}_ {\mathrm{i}\nu}(\tilde
  k\vert\phi\vert)\right]\ ,
\end{equation}
where $\tilde k= \mathrm{i}k$ and is positive. Here again it is more
convenient to rewrite the general matter wave function in terms of
Hankel functions in order to impose the boundary condition. We then
rewrite the previous wave function as
\begin{equation}
\varphi_k(\alpha,\vert\phi\vert)=\sqrt{\vert\phi\vert}\left[d_1\mathrm{H}_{\mathrm{i}\nu}^{(1)}(\tilde
  k\vert\phi\vert)+d_2\mathrm{H}_ {\mathrm{i}\nu}^{(2)}(\tilde
  k\vert\phi\vert)\right]\ ,
\end{equation}
where $d_1,d_2$ are arbitrary constants to be fixed by the boundary
condition. It can be checked that
$\mathrm{H}_{\mathrm{i}\nu}^{(1)}(\tilde k \vert\phi\vert)$ diverges
for large $\nu$ because
\begin{equation}
\mathrm{H}_{\mathrm{i}\nu}^{(1)}(x)=\frac{\sqrt{2}}{\pi}(\nu^2+x^2)^{-\frac14}\exp\left[\mathrm{i}\sqrt{\nu^2+x^2}-\mathrm{i}\,\nu\,\arcsinh\left(\frac{\nu}{x}\right)\right]
\exp\left[\frac{\pi}{2}\nu-\mathrm{i}\frac{\pi}{4}\right]\times\left[\mathrm{const.}+\mathcal{O}((\nu^2+x^2)^{-\frac12})\right],\,\,
\nu,x>0,
\end{equation}
cf. \cite{Bateman}, equation 7.13.2(22). On the other hand, it can be
shown that $\mathrm{H}_{\mathrm{i}\nu}^{(2)}(\tilde k \vert\phi\vert)$
vanishes at large values of $\nu$. This can be shown by combining  the
following properties of the Hankel functions:
$\mathrm{H}_{\mathrm{i}\nu}^{(2)}(x)=\exp(-\pi\nu)\mathrm{H}_{-\mathrm{i}\nu}^{(2)}(x)$,
$\mathrm{H}_{-\mathrm{i}\nu}^{(2)}(x)=(\mathrm{H}_{\mathrm{i}\nu}^{(1)}(x))^\star$,
where ${}^*$
denotes complex conjugation,  and the asymptotic expansion of
$\mathrm{H}_{\mathrm{i}\nu}^{(1)}(x)$ for large order as shown
above. Therefore, the wave function that fulfils the boundary
condition for negative energies reads
\begin{equation}
\varphi_k(\alpha,\vert\phi\vert)=d_2\sqrt{\vert\phi\vert}\mathrm{H}_ {\mathrm{i}\nu}^{(2)}(\tilde k\vert\phi\vert).
\label{negativeEphantom}
\end{equation}

Before proceeding further, it is worthwhile to point out the
following. The ordinary scalar field is defined on the entire real
line. The phantom field, on the other hand, is restricted to the
interval
$\vert\phi\vert\in\left[0;\phi_\star\right]$. Orthogonality
does therefore not hold straight away but follows only if equation
(\ref{orthogonality}) is satisfied in the boundaries in which $\phi$
is defined. To obtain orthogonality, additional conditions would be
necessary. As we cannot require that $\varphi_k$ vanishes at
$\phi_\star$, as this would be in conflict with the requirement that
the wave function follows the classical trajectory, we have to demand
that its derivative with respect to $\phi$ vanish there. This would
require an analytic expression of the zeros of the first derivative of
the modified Bessel function $\mathrm{K}_{i\nu}$ and the second Hankel function. To our knowledge, such an expression does
not exist. Also, there is no {\it physical} motivation for this
additional condition.

The $C_k$ have to satisfy
$\frac{\kappa^2}{6}\ddot{C}_k+k^2C_k=0$. The solution is given by Eq.~(\ref{gravsolution}): (i) For
positive energy, the imposition of the boundary condition leaves
$b_1$ and $b_2$ as arbitrary constants and therefore we need to
impose the boundary condition on the matter-dependent part in this
case, see Eq.~(\ref{positiveEphantom}). (ii) For negative energy, the
imposition of the boundary condition 
picks up the exponentially decreasing function.
The decay of the wave
function for large $\alpha$ is thus already guaranteed through the
purely gravitational part of the wave function. In principle, there is
no need to 
impose the boundary condition on the matter-dependent part in this
case, as long as the general solution for the matter part is
finite. However,  it is not the case, so we have as well to impose the
boundary condition on the matter part. This implies that only the
second kind of Hankel function $H_{i\nu}^{(2)}$ is present on the
matter part (see Eq.~(\ref{negativeEphantom})). 

In summary, the physical solutions are
\begin{eqnarray}
\Psi_k(\alpha,\phi)&=&\left(b_1e^{\mathrm{i}\frac{\sqrt6k}{\kappa}\alpha}+b_2e^{-\mathrm{i}\frac{\sqrt6k}{\kappa}\alpha}\right)\, 
\sqrt{\vert\phi\vert}\mathrm{K}_{i\nu}(k\vert\phi\vert),\,\,\, k^2>0\,
\nonumber\\ 
\Psi_k(\alpha,\phi)&=&d_2\exp\left(-\frac{\sqrt{6}}{\kappa}{\tilde
    k\alpha}\right)\sqrt{\vert\phi\vert}\mathrm{H}_
{\mathrm{i}\nu}^{(2)}(\tilde k\vert\phi\vert), \,\,\,  \,\,\, \,\,\,
\,\,\, \,\,\, \,\,\,k^2<0. 
\label{psik3}
\end{eqnarray}
These restrictions have to be taken into account when constructing wave
packets (which will not be attempted here). Before concluding this
subsection, we remark that the functions $\Psi_k(\alpha,\phi)$ given in
(\ref{psik3}) fulfil the DeWitt criterium as well, since they
approach zero for $\alpha=0$ and $\phi=0$.

\subsection{Smoothing the big-d\'emarrage singularity}\label{BCbd}

As a third model with the same quantum structure, we consider the
big-d\'emarrage singularity. The equations for this singularity as
it occurs for $\beta\gg1$ are just the same as for the big-freeze
generated by a phantom GCG presented above. The solution is also
given by Eq.~(\ref{generalvarphik}). The difference lies in the
classically forbidden region and thus in the boundary condition to
be employed. Whereas in the previously studied phantom model we
found a future singularity at $a=a_{\rm{max}}$, this model has a
past sudden singularity at $a=a_{\rm{min}}=1$. We therefore require
the wave function to satisfy $\Psi\to0$ as $\alpha\to-\infty$. But
as $\alpha\to-\infty$, $\nu\to\frac12$.

We split our analysis for positive and negative energies. For $k^2>0$,
the boundary condition on the matter part of the wave function is
$c_1=0$, that is, we are left with the first Bessel function, while the
gravitational part will be oscillatory. On the other hand, for $k^2<0$,
the boundary condition on the gravitational part of wave function
implies $b_1=0$, while  no other boundary condition has to be imposed on
the matter sector of the wave function. In summary, we obtain
\begin{eqnarray}
\Psi_k(\alpha,\phi)&=&\left(b_1e^{\mathrm{i}\frac{\sqrt6k}{\kappa}\alpha}+b_2e^{-\mathrm{i}\frac{\sqrt6k}{\kappa}\alpha}\right)\, \sqrt{\vert\phi\vert}\mathrm{J}_{\nu}(ik\vert\phi\vert),\,\,\, \,\,\, \,\,\, \,\,\,\,\, \,\,\,\,\ k^2>0\, \nonumber\\
\Psi_k(\alpha,\phi)&=&\exp\left(\frac{\sqrt{6}}{\kappa}k\alpha\right)\sqrt{\vert\phi\vert}\left[c_1\mathrm{J}_\nu(\tilde{k}\vert\phi\vert)+c_2\mathrm{Y}_\nu(\tilde{k}\vert\phi\vert)\right], 
\,\,\, \,\,\, \,\,\, \,\,\, \,\,\,k^2<0. 
\label{psik4}
\end{eqnarray}
where $\tilde k= \mathrm{i}k$.

The solution is thus similar to the one found for the standard scalar
GCG presented in the previous subsection. But note that $\phi$ is
restricted to a finite interval here.

Furthermore, as a consequence of the fact that the phantom scalar
field has compact range, the matter
dependent part of the wave function, $\varphi_k$, is not orthogonal for
different $k$. To obtain this, one has to require the condition
(\ref{orthogonality}). 

Here as well we notice that the functions $\Psi_k(\alpha,\phi)$ given in
(\ref{psik4}) fulfil the DeWitt criterium as they
vanish at $\alpha=0$ and $\phi=0$.


\subsection{Other boundary conditions}\label{BCother}

The Schr\"odinger equation with inverse square potential has been
studied by various authors \cite{Case:1950an, Frank:1971xx}.
They obtain as a solution the Hankel function of the first kind,
$\mathrm{H}^{(1)}_\nu$ for $k^2<0$ and $\tilde{V}_{\alpha}>\frac14$,
cf. Equation (3.6) in \cite{Frank:1971xx}.
Therefore, their solution is different from the ones we have
obtained in the previous subsections.
This is due to the boundary condition that is imposed in quantum
mechanics: $\Psi\to0$ as $r\to\infty$, $r$ being the radial
coordinate in the Schr\"odinger equation. In the quantum cosmological
model this would correspond to the condition $\Psi\to0$ as
$\vert\phi\vert\to\infty$. Whereas the vanishing of the wave
function at infinity is a sensible requirement in quantum mechanics,
it is not well-motivated in quantum cosmology. To demand a
square-integrable wave function makes sense only in a Hilbert space
with a probability interpretation and a (time) conserved
probability. It is not obvious that we have such a structure in
quantum cosmology: the Wheeler--DeWitt equation is timeless.

Applying nevertheless
this boundary condition to the quantum cosmological model
with GCG mimicked by a standard scalar field\footnote{Notice that
the boundary condition imposed in \cite{Case:1950an, Frank:1971xx}
cannot be applied in our model with a phantom scalar field. The
reason is that the scalar field is bounded in this case, so we
cannot take the limit $\vert\phi\vert\rightarrow\infty$ in
$\varphi_k(\alpha,\phi)$.}, we find
$\varphi_k(\alpha,\phi)=c_1\sqrt{\vert\phi\vert}\mathrm{H}^{(1)}_\nu(\mathrm{i}\tilde
k\vert\phi\vert)$ with $\tilde{k}^2=-k^2$  and $\nu$ purely
imaginary as $\tilde{V}_{\alpha}>\frac14$. This wave function
corresponds to the choice $c_2={\rm i}c_1$. Requiring in addition
orthogonality as in \cite{Frank:1971xx}, we would arrive at an
energy spectrum given by
 \be
\label{spectrum}
E_n(\alpha)=-\frac{\hbar^2\kappa_0^2}{2}\exp\left( \frac{2\pi
n}{\sqrt{\tilde{V}_{\alpha}-\frac14}}\right)\, , \quad n\in
\mathbb{Z} \ee where $\kappa_0$ is an {\em arbitrary} real constant. This
can only be done for $k^2<0$. This is a similar non-uniqueness than in
the quantum mechanical case \cite{Frank:1971xx}. 

Now, we can impose two different types of boundary condition on the
total wave function $\Psi$:  

\begin{enumerate}

\item $C_k$ vanishes at the singularity; i.e. at $\alpha\approx0$,
  and $\varphi_k(\alpha,\phi)$ is bounded at  
$\alpha\approx0$ and $\phi\approx0$, or 

\item $C_k$ vanishes well inside the forbidden region;
  i.e. $\alpha\to-\infty$, and $\varphi_k(\alpha,\phi)$ is bounded for
  $\alpha\to-\infty$ and $\phi\approx0$. We next impose both
  conditions separately. 

\end{enumerate}

The equation for the gravitational part of general solution for the
wave function obtained through a Born--Oppenheimer approximation
(see Eq.~(\ref{Ckeqn})) can then be solved in the vicinity of the
singularity, i.e. for $\alpha\approx0$. It satisfies
\begin{equation}
\ddot C_k- \gamma\left(1-\delta\alpha\right)C_k=0,
\end{equation}
where the constants  $\gamma,\delta$ and $c$ read
\begin{equation}
\gamma= \frac{6\kappa_0^2}{\kappa^2}\exp\left[\frac{4\pi
n}{\sqrt{4c-1}}\right], \quad \delta= \frac{48\pi n
c}{(4c-1)^\frac32}, \quad c=\frac2{\hbar^2}a_0^6
V_\ell\left[\frac{\sqrt{3}}{2}\kappa\vert 1+\beta\vert\right]^{-2}.
\end{equation}
Therefore, the gravitational part of the wave function for
$\alpha\approx0$ reads \cite{Abramowitz}
\begin{equation}
C_k(\alpha)=b_1
\mathrm{Ai}\left[\left(\frac{\gamma}{\delta^2}\right)^\frac13-\left(\delta\gamma\right)^\frac13\alpha\right]+b_2
\mathrm{Bi}\left[\left(\frac{\gamma}{\delta^2}\right)^\frac13-\left(\delta\gamma\right)^\frac13\alpha\right],
\label{obcg1}
\end{equation}
where $\mathrm{Ai}(z)$, $\mathrm{Bi}(z)$ denote Airy functions. Imposing that the wave function $C_k$  vanishes at the singularity ($\alpha=0$) implies 
\begin{equation}
 b_2=-b_1 \frac{\mathrm{Ai}\left[\left(\frac{\gamma}{\delta^2}\right)^\frac13\right]}{\mathrm{Bi}\left[\left(\frac{\gamma}{\delta^2}\right)^\frac13\right]}.
\end{equation}
It can be proven as well that $\varphi_k(\alpha,\phi)=c_1\sqrt{\vert\phi\vert}\mathrm{H}^{(1)}_\nu(\mathrm{i}\tilde
k\vert\phi\vert)$ vanishes for  $\alpha\approx 0$ and $\phi \approx
0$. Therefore, 
we can conclude that the DeWitt criterium is compatible (in this case)
with the boundary condition used in \cite{Case:1950an, Frank:1971xx}. 

The boundary condition $\Psi\to0$ as $\alpha\to-\infty$ cannot be imposed on
the previous wave function (\ref{obcg1}). This is so because the wave
function  (\ref{obcg1}) is valid only around
$\alpha\approx0$. In order to impose the boundary condition $\Psi\to0$
as $\alpha\to-\infty$,  we have to consider
the wave function  for $\alpha\to-\infty$.
In this limit, the energy spectrum (\ref{spectrum})
reduces to\footnote{In taking this limit, notice that $E_n(\alpha)$ can become complex. However, $E_n(\alpha)$ given in (\ref{Enlimit}) is always positive.}
\be
E_n(\alpha)=-\frac{\hbar^2\kappa_0^2}{2}\, . 
\label{Enlimit}
\ee 
Consequently, \be
C_k(\alpha)=b_1\exp\left(\frac{\sqrt{6}\kappa_0}{\kappa}\alpha\right)+b_2\exp\left(-\frac{\sqrt{6}\kappa_0}{\kappa}\alpha\right). \ee
Now, the condition $C_k\to 0$ as $\alpha\to-\infty$ implies
$b_2=0$. On the other hand,  
$\varphi_k(\alpha,\phi)=c_1\sqrt{\vert\phi\vert}\mathrm{H}^{(1)}_\nu(\mathrm{i}\tilde 
k\vert\phi\vert)$ is bounded for $\alpha\to-\infty$ and $\phi \approx
0$. Therefore, we can conclude that the condition  $\Psi\to0$ as
$\alpha\to-\infty$ is fulfilled and compatible with the boundary
condition used in \cite{Case:1950an, Frank:1971xx}. 

In summary, as the wave function  remains finite, the singularity
would be avoided 
with the boundary conditions 1. and 2. as well. But we emphasize again that
the condition $\Psi\to0$ as $\vert\phi\vert\to\infty$ is perhaps not
obligatory in the quantum cosmological case. It is presented here to
compare our results to results presented elsewhere in the
literature.


\section{Conclusions and outlook}
\label{end}

What are the main results of our paper? We have shown how the
classical singularities of particular cosmological models describing dark energy features
 can be avoided in quantum cosmology. The framework is quantum
geometrodynamics, and restriction has been made to a class of models
(containing a generalized Chaplygin gas) which could be of relevance
for the description of dark energy. The wave function $\Psi$ is
defined on a two-dimensional configuration space consisting of the
scale factor $a$ and the scalar field $\phi$ representing the gas.
After employing a Born--Oppenheimer type of approximation, we arrive
at an effective stationary Schr\"odinger equation with singular (in
the sense of quantum mechanics) potentials. The occurrence of such
potentials in quantum mechanics signals an essential non-uniqueness
of the solutions. The same happens here. Requiring that the wave
function go to zero in the classically forbidden region still leaves
a whole class of solutions which encode singularity avoidance in the
sense of vanishing wave function (DeWitt-criterium). This is
different from our earlier papers \cite{DKS} and
\cite{Kamenshchik:2007zj} in which we dealt with ``regular''
potentials and where appropriate normalizability assumptions singled
out singular-free solutions (cf. also \cite{Barboza:2006an}). The
case of the big-brake singularity in \cite{Kamenshchik:2007zj}
contains a potential that behaves like the Coulomb potential in the
vicinity of the classical singularity; as in quantum mechanics, no
ambiguity remains after normalizability is being assumed.

Our analysis has been performed in detail only for the case of the
$r^{-2}$-potential. A similar discussion can be made for the
$r^{-4}$-potential leading to Mathieu functions; since the results are
analogous to those of the $r^{-2}$-potential, we do not present them
here. We expect that a similar, although much more complicated,
analysis can be performed for other singular potentials as well.
It would be desirable to have a general classification of
singularities with respect to singularity avoiding states at hand.

We want to emphasize again one important conceptual aspect of this
(and earlier) work. The singularities in these models do generally
occur for a big universe, that is, for a value of the scale factor
much bigger than the Planck scale. Singularity avoidance by quantum
gravity then necessarily entails quantum effects for a big universe.
Such quantum effects have been discussed earlier in the case of a
classically recollapsing quantum universe \cite{KZ95}; the quantum
effects occur there near the
(non-singular) region of the classical turning point
as a consequence of the boundary conditions. We thus see that quantum
cosmology may be of direct relevance for the future of our Universe.
As discussed at length in \cite{KZ95}, the occurrence of such quantum
effects possesses essential importance for understanding the arrow of
time. It would be of great interest to study the question of entropy
increase in the present models.

There are many further open questions. In quantum mechanics,
the remaining ambiguity of the solutions and the ensuing spectra
points to new physics a short distances. The ambiguity must therefore
be fixed either by reference to experimental results or by knowledge
of the deeper theory. How is the situation in quantum cosmology?
Experiments do not yet seem available, but can the ambiguity be fixed
by theoretical means? One possible project would be to investigate
whether prominent boundary conditions such as the no-boundary
\cite{Hartle:1983ai} or tunnelling \cite{Vilenkin} conditions are able to
fix it. Perhaps the ambiguity can only be avoided by going to a more
fundamental theory such as string theory. On the other hand, it may as
well be imaginable that cosmological models leading to such singular
matter potentials have to be excluded---an interesting selection
criterium for equations of state.

An essential ingredient in our whole analysis is the demand that the
wave function go to zero in classically forbidden regions. How this
relates in general to the hyperbolic structure of the
Wheeler--DeWitt equation is not fully understood. Here we have
avoided this problem by going to the Born--Oppenheimer
approximation. In general one would expect that certain quantization
conditions on the physical parameters (masses, coupling constants,
cosmological constant) may appear
\cite{CK90,Gerhardt,BouhmadiLopez:2004mp,BouhmadiLopez:2006pf}. It
would also be very interesting to establish a connection with a
group-theoretical generalization of quantum geometrodynamics where
it was shown that not only time, but even the ordinary
three-dimensional space is absent at the most fundamental level,
leading to singularity avoidance in a natural way
\cite{DamourNicolai}. We hope that we can address some of these open
issues in future publications.

Finally, let us point out  that a future sudden singularity has been
recently investigated in \cite{Cailleteau:2008wu}. The analysis was
carried out in the loop quantum cosmology (LQC) framework\footnote{A
  discussion on the standard big bang singularity and LQC can be found in reference \cite{Bojowald:2007ky}.}. In
this model, 
matter is modelled by a scalar field which rolls down (standard scalar
field) or up (phantom scalar field)  through an 
unbounded exponential potential. In the framework of LQC this  scalar field behaves such that its effective energy 
density is finite; that is, there is a balance between its kinetic
energy and its potential, while the effective pressure blows up in a
finite future cosmic time. It was then concluded in 
\cite{Cailleteau:2008wu} that such a singularity cannot be avoided
by means of the effective Friedmann equation in LQC. As it is mentioned in \cite{Cailleteau:2008wu},  the sudden singularity that appears for the phantom scalar field in the context of the modified Friedmann equation in LQC corresponds to a big-rip singularity in the standard relativistic case. For such a big-rip singularity, it was shown in \cite{DKS} that it can be avoided in the sense that wave packets necessarily disperse when they approach the region of the classical big-rip singularity. It would be interesting to see if this sort of big-rip singularity can be avoided as well by employing the DeWitt criterium that the wave function be zero at the classical singularity.


\section*{Acknowledgments}
MBL is  supported by the Portuguese Agency Funda\c{c}\~{a}o para a Ci\^{e}ncia e
Tecnologia through the fellowship SFRH/BPD/26542/2006. She also wishes
to acknowledge the hospitality of the Institute of
Theoretical Physics of the University of Cologne during the completion
of part of this work.
 This research work was
supported by the grant FEDER-POCI/P/FIS/57547/2004. CK acknowledges
kind hospitality at the Max Planck Institute for Gravitational
Physics, Potsdam, Germany, while part of this work was done.

\appendix

\section{Justification of the gravitational wave function approximation}
\label{justBO}

In our models, we constructed the potential $V(\phi)$ such that it
corresponds to  the polytropic equation of state for a GCG. The
potential therefore depends on $\kappa$ and this dependence carries
over to the matter-dependent part of the wave function.
The Born--Oppenheimer approximation can be understood as an expansion
scheme with respect to $\kappa$.

The derivatives of $\varphi_k$ with respect to $\alpha$ are
of non-zero order in $\kappa$. This comes from
$\dot\varphi_k=\frac{\d\varphi_k}{\d\nu^2}\frac{\d\nu^2}{\d\alpha}$.
In the vicinity of the singularity, $\alpha\approx0$ and
$\frac{\d\nu^2}{\d\alpha}=-\frac{16}{\hbar^2}\frac{a_0^6}{\kappa^2}\frac{V_\ell}{\vert1+\beta\vert^2}$.
Recall that $V_\ell=\frac{\vert A\vert^{\frac{1}{1+\beta}}}{2}$ and
$a_0=\left|\frac{B}{A}\right|^{\frac{1}{3(1+\beta)}}$.
  Recall also  that we use $\nu_0$ to denote the value of $\nu$ at
  $\alpha=0$. Therefore
\be
\frac{{\rm d}\nu^2}{{\rm d}\alpha}=-\frac{8}{\hbar^2\kappa^2}\left|\frac{B^2}{A}\right|^{\frac{1}{1+\beta}}\frac{1}{\vert1+\beta\vert^2}.
\ee
 From Eq.~(\ref{gravsolution}) we see that
$C_k\sim\mathcal{O}\left(\kappa^0\right)$,
$\dot{C}_k\sim\mathcal{O}\left(\kappa^{-1}\right)$ and
$\ddot{C}_k\sim\mathcal{O}\left(\kappa^{-2}\right)$.

To obtain a consistent approximation scheme, such that in  \be
\ddot{C}_k\varphi_k\sim\mathcal{O}\left(\kappa^{-2}\right)\, \qquad
\dot{C}_k\dot\varphi_k\sim\mathcal{O}\left(\kappa^{-1}
\right)\ ,\qquad
C_k\ddot\varphi_k\sim\mathcal{O}\left(\kappa^{1}\right)\ , \ee only
the first term is kept and the others neglected, we have to assume
that $\left|\frac{B^2}{A}\right|^{\frac{1}{1+\beta}}\sim\kappa^2$,
i.e.,
$\frac{\d\nu^2}{\d\alpha}\sim\mathcal{O}\left(\kappa^0\right)$.


\end{document}